\documentclass[amsmath,amssymb,superscriptaddress,nobalancelastpage,aps,prl,twocolumn]{revtex4-2}

\usepackage{graphicx}
\usepackage{varioref}
\usepackage{xr-hyper}
\usepackage{xcolor}
\usepackage{nicefrac}
\usepackage{xfrac}
\usepackage{hyperref}
\hypersetup{colorlinks,linkcolor=blue,urlcolor=blue,citecolor=blue}
\usepackage{ulem}
\usepackage{lineno}
\usepackage{amsmath} 
\usepackage{amssymb}
\usepackage{siunitx}

\newcommand{\LSCO}{La$_{1.875}$Sr$_{0.125}$CuO$_4$}

\begin{document}

\title{Fate of charge order in overdoped La-based cuprates}

%-----------------------------------
\author{\underline{K.~von~Arx}}
\email{karin.vonarx@uzh.ch }
\affiliation{Physik-Institut, Universit\"{a}t Z\"{u}rich, Winterthurerstrasse 
190, CH-8057 Z\"{u}rich, Switzerland}
\affiliation{Department of Physics, Chalmers University of Technology, SE-412 96 G\"{o}teborg, Sweden}

\author{Qisi~Wang}
\affiliation{Physik-Institut, Universit\"{a}t Z\"{u}rich, Winterthurerstrasse 
190, CH-8057 Z\"{u}rich, Switzerland}

\author{S. Mustafi}
\affiliation{Physik-Institut, Universit\"{a}t Z\"{u}rich, Winterthurerstrasse 190, CH-8057 Z\"{u}rich, Switzerland}

\author{D.~G.~Mazzone}
\affiliation{Laboratory for Neutron Scattering and Imaging, Paul Scherrer Institut, CH-5232 Villigen PSI, Switzerland}

\author{M.~Horio}
\affiliation{Institute for Solid State Physics, The University of Tokyo, Kashiwa, Chiba 277-8581, Japan}

\author{D.~John~Mukkattukavil}
\affiliation{Department of Physics and Astronomy, Uppsala University, SE-75121 Uppsala, Sweden}

\author{E.~Pomjakushina}
\affiliation{Paul Scherrer Institut, CH-5232 Villigen PSI, Switzerland}

\author{S.~Pyon}
\affiliation{Department of Applied Physics, The University of Tokyo, Tokyo 113-8646, Japan}

\author{T.~Takayama}
\affiliation{Max Planck Institute for Solid State Research, 70569 Stuttgart, Germany}

\author{H.~Takagi}
\affiliation{Max Planck Institute for Solid State Research, 70569 Stuttgart, Germany}
\affiliation{Department of Physics, The University of Tokyo, Tokyo 113-0033, Japan}

\author{T.~Kurosawa}
\affiliation{Department of Physics, Hokkaido University, Sapporo 060-0810, Japan}

\author{N.~Momono}
\affiliation{Department of Physics, Hokkaido University, Sapporo 060-0810, Japan}
\affiliation{Department of Applied Sciences, Muroran Institute of Technology, Muroran 050-8585, Japan}

\author{M.~Oda}
\affiliation{Department of Physics, Hokkaido University, Sapporo 060-0810, Japan}

\author{N.~B.~Brookes}
\affiliation{European Synchrotron Radiation Facility, B.P. 220, 38043 Grenoble, France}

\author{D.~Betto}
\affiliation{European Synchrotron Radiation Facility, B.P. 220, 38043 Grenoble, France}

\author{W.~Zhang}
\affiliation{Swiss Light Source, Photon Science Division, Paul Scherrer Institut, CH-5232 Villigen PSI, Switzerland}

\author{T.~C.~Asmara}
\affiliation{Swiss Light Source, Photon Science Division, Paul Scherrer Institut, CH-5232 Villigen PSI, Switzerland}

\author{Y.~Tseng}
\affiliation{Swiss Light Source, Photon Science Division, Paul Scherrer Institut, CH-5232 Villigen PSI, Switzerland}

\author{T.~Schmitt}
\affiliation{Swiss Light Source, Photon Science Division, Paul Scherrer Institut, CH-5232 Villigen PSI, Switzerland}

\author{Y.~Sassa}
\affiliation{Department of Physics, Chalmers University of Technology, SE-412 96 G\"{o}teborg, Sweden}

\author{J.~Chang}
%\email{johan.chang@physik.uzh.ch}
\affiliation{Physik-Institut, Universit\"{a}t Z\"{u}rich, Winterthurerstrasse 190, CH-8057 Z\"{u}rich, Switzerland}

%-----------------------------------

\maketitle
\textbf{In high-temperature cuprate superconductors, stripe order refers broadly to a coupled spin and charge modulation with a commensuration of eight and four lattice units, respectively. How this stripe order evolves across optimal doping remains a controversial question. Here we present a systematic resonant inelastic x-ray scattering (RIXS) study of weak charge correlations in La$_{2-x}$Sr$_x$CuO$_4$ (LSCO) and La$_{1.8-x}$Eu$_{0.2}$Sr$_x$CuO$_4$ (LESCO).
Ultra high energy resolution experiments demonstrate the importance of the separation of inelastic and elastic scattering processes. Upon increasing doping $x$, the long-range temperature dependent stripe order is found to be replaced by short-range temperature independent correlations at a critical point $x_c\approx 0.15$ distinct from the pseudogap critical doping. We argue that the doping and temperature independent short-range correlations originate from unresolved electron-phonon coupling that broadly peaks at the stripe ordering vector. In LSCO, long-range static stripe order vanishes in a quantum critical point around optimal doping.}
\\[2mm]

%\section{Introduction}
Charge order is now established in virtually all known hole underdoped cuprates~\cite{LaliberteNatComm2011,Wu11,chang12,GhiringhelliSCI12,TabisNatComm2014,tranquada,HuckerPRB2011,AchkarPRL2012,ThampyPRB2014,CominScience2014,AchkarSCI2016}, and hence is a universal property 
on equal footing with the pseudogap~\cite{Norman2015} and superconductivity. The evolution of charge order and the pseudogap  
beyond optimal doping has been challenging to establish. The pseudogap may emerge through a quantum critical transition~\cite{DaouNat2010,HashimotoNatMat2015,HorioPRL2018,MichonNature2019,gupta_vanishing_2021}, a cross-over phenomenon~\cite{TallonPRB2020} or as a precursor to a symmetry breaking. 
It has also been difficult to establish a general trend as to how charge order evolves into the overdoped limit.
In (Bi,Pb)$_{2.12}$Sr$_{1.88}$CuO$_{6+x}$ (Bi2201), a long-range charge ordering extending all the way to room temperature is reported in the overdoped regime beyond the doping extent of the pseudogap phase~\cite{PengNatMat2018}. 
In the La-based cuprates, spin and charge stripe order around 1/8-doping are coupled~\cite{tranquada,TranquadaPRB96,HuckerPRB2011,ChangPRB2008,LakeNature2002}.
Doping evolution of stripe order thus involves both spin and charge unless a doping-induced decoupling takes place. A recent study on La$_{1.6-x}$Nd$_{0.4}$Sr$_x$CuO$_4$ suggests that spin-stripe order persists across the entire superconducting dome~\cite{MaPRR2021} and long-range charge ordering outside the pseudogap phase has been reported in La$_{2-x}$Sr$_x$CuO$_4$ (LSCO)~\cite{LinPRL2020, MiaonpjQM2021}. A resonant x-ray scattering study, by contrast, suggests that short-range charge correlations only exist inside the pseudogap phase~\cite{WenNatComm2019}. Thus, contradicting observations on the charge order leave the questions of the coupling between spin and charge as well as a possible connection between charge order and the pseudogap unsolved.

\begin{figure*}[t]
 	\begin{center}
 		\includegraphics[width=0.99\textwidth]{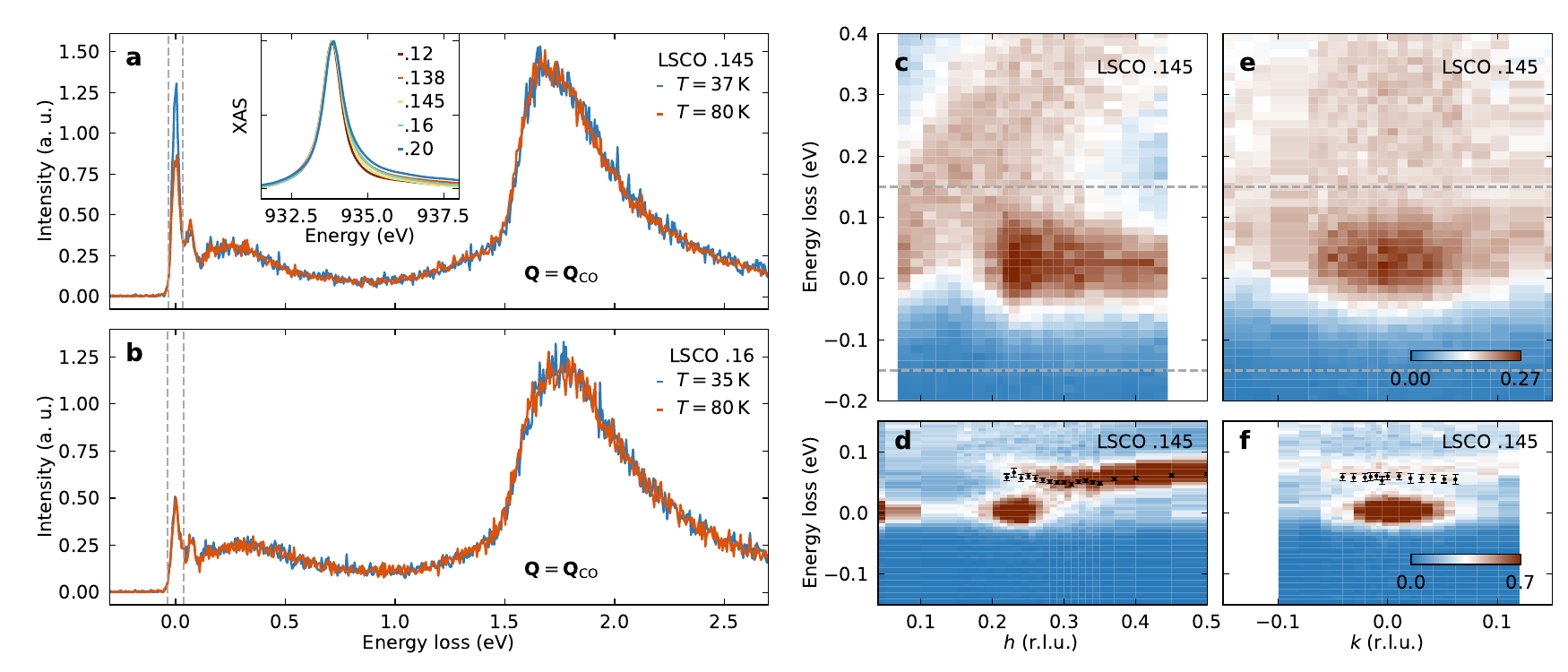}
 	\end{center}
 	\caption{
 	X-ray absorption and resonant inelastic x-ray scattering
 	spectra recorded on \LSCO\ as a function of energy loss and momentum. (a), (b) RIXS spectra recorded on  LSCO $x=0.145$ and 0.16  at the charge ordering vector for different temperatures as indicated. Vertical dashed lines indicate the energy resolution and with that the integration window of elastic scattering. The intensity is given in arbitrary units (a. u.). The inset shows XAS spectra featuring the copper $L$-edge for LSCO with doping concentrations as indicated. (c-f) RIXS spectra probed in longitudinal ($h$) and transverse ($k$) directions on LSCO $x=0.145$.
 	Data in (c,e) are recorded with an energy resolution of 129 meV, whereas (d,f) show spectra recorded at a high resolution beamline with a total resolution of 33 meV. Horizontal dashed lines in (c,e) illustrate the energy range in (d,f). The improved resolution allows for resolving the phonon branch. The black circles mark the phonon dispersion determined from fitting the spectra.
 	High resolution data were taken at 37~K, all other data at base temperature, see methods section.}
	\label{fig:fig1}
 \end{figure*}
 
Recently, Resonant Inelastic X-ray Scattering (RIXS) has been used to detect weak charge correlations by separating elastic from inelastic contributions~\cite{WangPRL2020}. In this fashion, charge correlations in La$_{1.8-x}$Eu$_{0.2}$Sr$_x$CuO$_4$ (LESCO) $x=0.12$ were probed above the resonant elastic x-ray scattering onset temperature~\cite{FinkPRB2009,FinkPRB2011,AchkarSCI2016}.
Here we use the RIXS sensitivity to trace charge correlations as a function of doping and temperature in LSCO and LESCO. 
Our main finding is the presence of long-range charge order for $x<0.15$ and short-range correlations for $x>0.15$.
We show how the long-range charge order is strongly temperature and doping dependent while the short-range correlations are essentially doping and temperature independent.
Our study provides a complete charge order phase diagram across optimal doping. Long-range charge correlation vanishes around optimal doping. As the correlation length is vanishing a quantum critical scenario is conceivable. Short-range charge correlations are found beyond optimal doping and seemingly beyond the pseudogap phase. We thus conclude that neither the long- nor short-range charge correlations correlate with the pseudogap phenomenon.

\begin{figure*}[t]
 	\begin{center}
 		\includegraphics[width=0.99\textwidth]{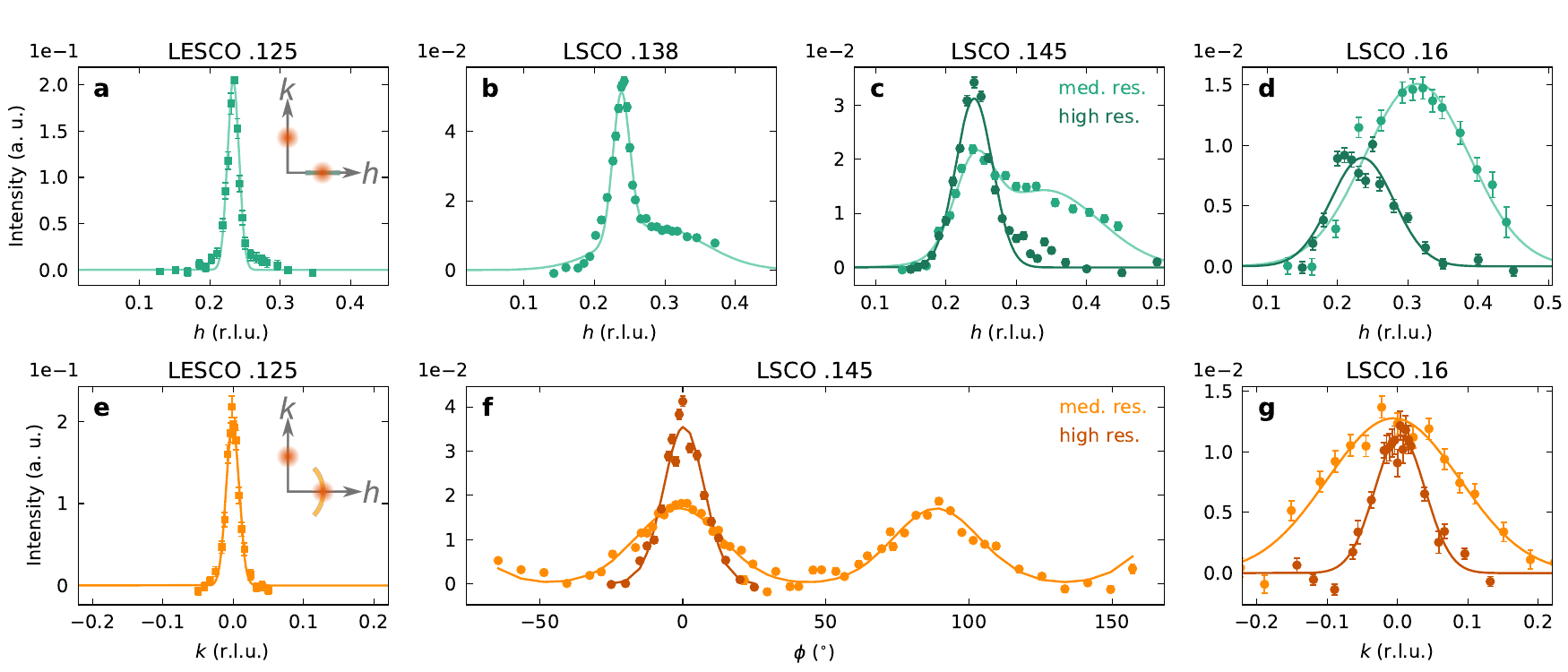}
 	\end{center}
 	\caption{ Doping evolution of the charge order in LSCO and LESCO. (a-d)  Longitudinal scans along $(h,0)$ for the compounds and compositions as indicated.	(e,g) Transverse scans through the longitudinal peak position $(\delta,k)$ for doping concentrations as indicated. (f) Circular arc scan through two charge order reflections, with $\phi=0$ at ($h$,0). Panels for LSCO $x=0.145$ and 0.16 compare high- and medium- resolution data. Solid lines are Gaussian fits. See text for further explanations. Insets display sketches of the charge order reflections in reciprocal space and the scan trajectories.  Data in  (a,e) are adapted from Ref.~\cite{WangPRL2020}.}
	 \label{fig:fig2}
\end{figure*}

\noindent\textbf{Results}\\
\textbf{Charge order correlations.} X-ray absorption spectra (XAS) across the copper $L$-edge for different doping concentrations $x$ of LSCO are shown in the inset of Fig.~\ref{fig:fig1}a. Comparable data quality is obtained irrespective of doping and consistent with previous XAS studies of LSCO~\cite{BrookesPRL2015}. Example RIXS spectra on LSCO $x=0.145$ and $x=0.16$, recorded at the charge ordering wave vector for two different temperatures
%\sout{and a background position} 
reveal scattering from elastic processes,  phonon (70 meV), spin  ($\sim0.3$ eV) and $dd$ (1.8 eV) excitations (Fig.~\ref{fig:fig1}a,b).The phonon-, spin- and $dd$- are consistent with previous RIXS studies~\cite{LinPRL2020, WangSciAdv2021}. A comparison of the two raw spectra shows that the elastic intensity is significantly enhanced at low temperature for $x=0.145$ whereas elastic scattering appears temperature independent for $x=0.16$.
The momentum dependence of elastic and inelastic scattering recorded on LSCO $x=0.145$ is shown in Fig.~\ref{fig:fig1}c-f for longitudinal ($h$) and projected transverse ($k$) directions. 
In Fig.~\ref{fig:fig1}c,e data are recorded with an overall energy resolution of 129~meV (FWHM), whereas in d,f the resolution is 33~meV. The direct comparison illustrates how the quasielastic scattering in c and e includes both elastic and phonon responses. Since the intensity from the charge order peak and the phonon branch are comparable, differentiation can only be reached by applying sufficiently high energy resolution. Lack of sufficient energy resolution implies a convolution of charge order and phonon scattering intensities. The optical phonon branch, extracted from the high resolution data of LSCO $x=0.145$ in Fig.~\ref{fig:fig1}d and of LSCO $x=0.16$ in supplementary Fig.~S3 show a dispersion behaviour consistent with previous studies~\cite{WangSciAdv2021, LinPRL2020}.

\begin{figure*}
    \begin{center}
 		\includegraphics[width=0.99\textwidth]{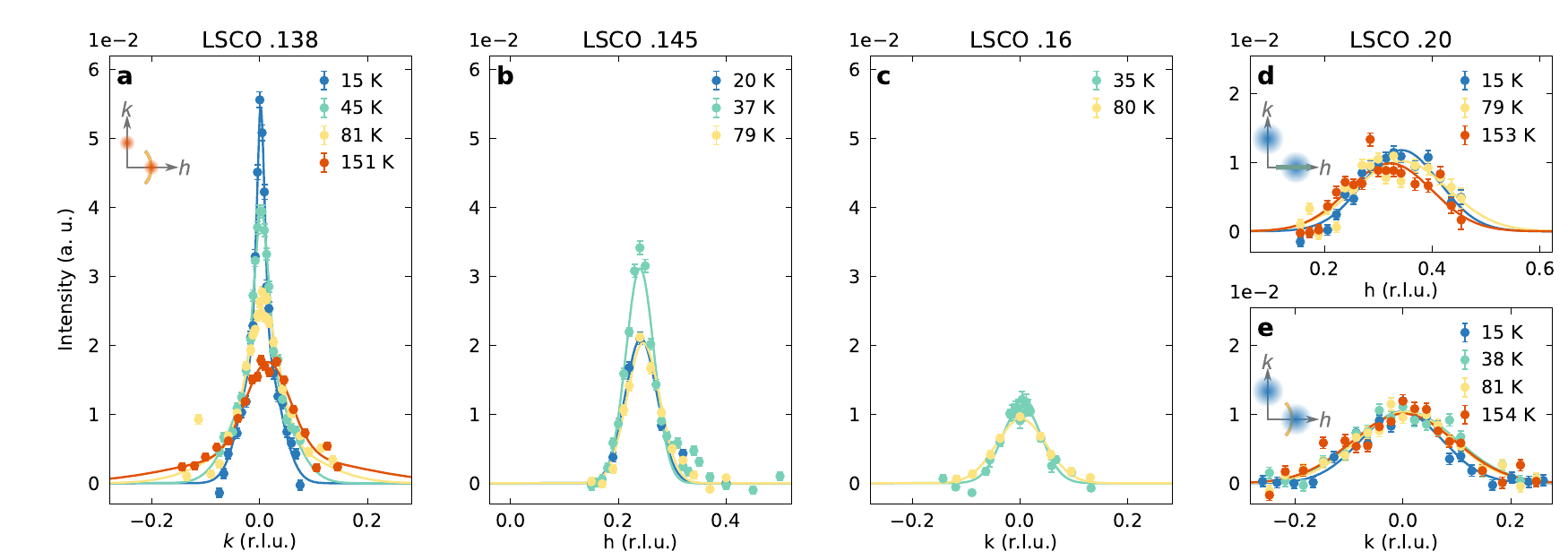}
    \end{center}
    \caption{Charge order temperature and doping dependence. 
    (a-e) Transverse and longitudinal scans through the charge ordering vector ($\delta,0$) recorded as a function of temperature, as indicated, on LSCO $x=0.138, 0.145, 0.16$ and $0.20$. Data in (b,c) were recorded with high energy resolution, % \sout{(33 meV FWHM)},
    %\sout{a 130 meV}
    whereas in (a,d,e) medium resolution was applied, see text for further information. Solid lines are fits using a Gaussian lineshape. All intensities have been normalized to the integrated $dd$-excitation intensity. Insets in (a,d,e) display the longitudinal and transverse scan direction.
 	}	
    	\label{fig:fig3}
\end{figure*}

In what follows, quasielastic intensity is defined by integrating the RIXS intensity in a window set by the respective energy resolution around zero energy loss -- see Fig.~\ref{fig:fig1}a.
In this fashion, longitudinal and transverse elastic scans through the charge order vector were carried out for LSCO ($x=0.12$, 0.138, 0.145, 0.16 and 0.2) and LESCO ($x=0.125$, and 0.21) crystals. As shown in Fig.~\ref{fig:fig2}a, stripe order in LESCO $x=0.125$ manifests by a reflection at $Q=(\delta,0)$ with $\delta=0.23$, as previously reported~\cite{WangPRL2020}. At optimal doping ($x\sim 0.16$ shown in Fig.~\ref{fig:fig2}d,g) and in the overdoped limit, represented by LSCO $x=0.20$ (Fig.~\ref{fig:fig3}d,e) and LESCO $x=0.21$ (supplementary Fig.~S2), much less elastic scattering is found. 

For longitudinal ($h$) scans recorded with medium resolution, a double peak structure is found (see Fig.~\ref{fig:fig2}b,c). The peak at $(h,0)\approx(1/3,0)$ stems from the optical phonon mode, likely in combination with absorption effects for large values of $h$. In transverse (rocking) scans,  illustrated in the inset of Fig.~\ref{fig:fig2}e, the charge order peak appears with additional broadening (see Fig.~\ref{fig:fig2}f,g).
Once sufficient energy resolution is applied, the double peak structure disappears in the longitudinal scans and sharper peaks are found in the transverse scans. This comparison shows the importance of inelastic experiments as a probe of weak charge order away from 1/8 doping. \\

\begin{figure*}[t]
    \begin{center}
		\includegraphics[width=0.99\textwidth]{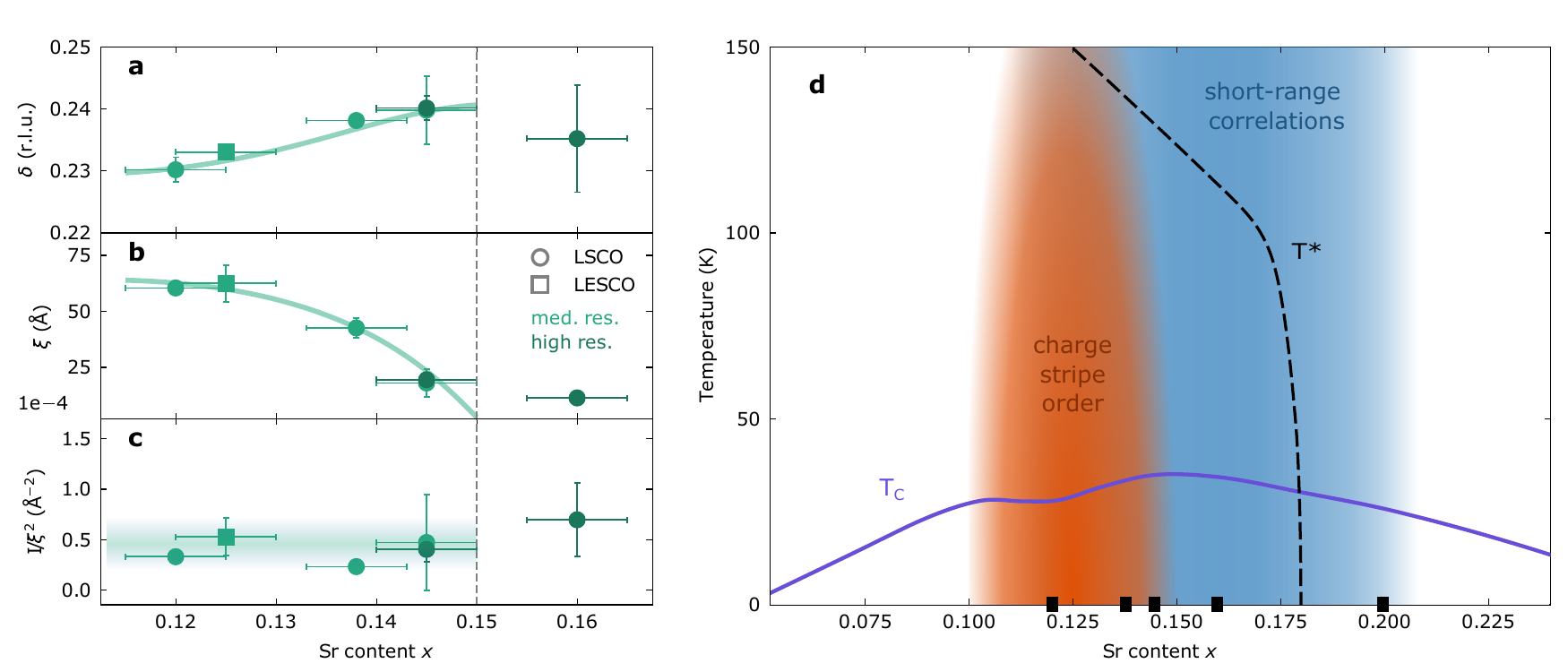}
    \end{center}
    \caption{(a-c) Charge order incommensurability $\delta$, correlation length $\xi$ (defined as the inverse half-width-half-maximum) and integrated diffracted intensity (amplitude $I$ divided by $\xi^2$ ) versus hole doping for LSCO (circles) and LESCO (squares) at base temperature.Error bars reflect the standard deviation obtained from fits to the diffraction data. Data for LESCO $x=0.125$ are from Ref.~\cite{WangPRL2020}. All colored lines are guides to the eye. Vertical dashed line in (a-c) marks the critical doping separating long- from short-range correlations. (d) Charge order phase diagram: temperature versus hole doping/Sr content $x$. Superconducting dome and pseudogap onset of LSCO are illustrated by respectively solid violet and dashed black line. Red and blue phases indicate respectively long-range and short-range charge correlations. Solid squares indicate the doping compositions studied in this work.}
    \label{fig:fig4}
\end{figure*}

\noindent\textbf{From long-range order to short-range correlations} In between 1/8 and optimal doping, our longitudinal scans show a decreasing elastic scattering intensity as doping is gradually increased. 
The evolution in doping shows a distinct transition in the nature of the charge correlation when the temperature dependence is taken into account. In Fig.~\ref{fig:fig3}a-e, we display the temperature dependence %\sout{of longitudinal scan} 
for the LSCO $x=0.138$, 0.145, 0.16 and 0.20 compounds. For $x>0.15$, essentially no temperature dependence is observed up to 150~K. We note that these two compounds have been recorded with different energy resolution. However, the direct comparison for the same compound in Fig.~\ref{fig:fig1}c-f and Fig.~\ref{fig:fig2}c demonstrates how the phonon contribution is altering the elastic scans for the medium-resolution resolution data. Since the RIXS phonon intensity shows essentially no doping dependence for LSCO $x=0.12-0.21$~\cite{LinPRL2020}, the almost identical medium-resolution scans for $x=0.16$ and 0.20 lead to the conclusion that the elastic scattering must behave similarly. By contrast, for LSCO $x=0.138$  and 0.145 a pronounced temperature dependence is found. The charge order intensity is suppressed with increasing temperature for $T>T_c$. The long-range charge order is also partially suppressed for $T<T_c$ due to phase competition with superconductivity~\cite{ChangPRB2008}. By contrast, the short-range charge correlations appear insensitive to phase competition with superconductivity (Fig.~\ref{fig:fig3}c-e).
This imposes a picture of two types of charge order: one long-range order that decays with temperature and short-range temperature independent correlations. Our study provides a very narrow doping sector for the transition from temperature dependent long-range order to temperature independent short-range correlations between 0.145 and 0.16 doping. In fact, the transition may happen through a quantum critical point at $x\approx 0.15$. In Fig.~\ref{fig:fig4}a-c, we show how the incommensurability $\delta$, correlation length $\xi$ and integrated intensity $I/\xi^2$  evolve from 1/8 to the overdoped regime. While the integrated intensity $I/\xi^2$ is roughly independent of doping, the correlation length declines as $x\rightarrow 0.15$. At the same time, we observe that  $\delta \rightarrow 1/4 $ for $x\rightarrow 0.15$. Beyond $x=0.15$, the temperature independent charge correlations hold similar integrated weight and incommensurability. Their correlation length of 10~\AA\ corresponds to about three in-plane lattice parameters.\\

\noindent\textbf{Discussion}\\
Several contradicting studies of stripe order in overdoped La-based cuprates have been published recently~\cite{MiaonpjQM2021,WenNatComm2019,MaPRR2021,FrachetNatPhys2020,LinPRL2020,Lee_generic_2021}. At 1/8 doping, charge order in LESCO could be detected above the onset temperature of the low-temperature tetragonal (LTT) phase~\cite{KlaussPRL2000} and above the transport onset of the pseudogap~\cite{PGPRB2018}. As a function of doping, long-range charge order has been reported even outside the pseudogap phase of LSCO~\cite{LinPRL2020, MiaonpjQM2021}. Another study reported short-range charge correlations to exist within the pseudogap phase only and long-range
charge order only around 1/8-doping~\cite{WenNatComm2019}.
Our comprehensive inelastic study suggests short-range correlations for $x>0.15$ that persist across the pseudogap critical doping. The long-range stripe order by contrast decays fast as $x\rightarrow 0.15$. Our study therefore also contrasts the recent study reporting long-range charge order in LSCO, up to $x=0.21$~\cite{LinPRL2020}, despite using the same measurement technique. Neither of the short or long-range charge correlations reported here seem to connect with the psesudogap phase. A recent magneto-transport experiment across the low-temperature pseudogap ``critical'' point~\cite{fang2020fermi} suggests a symmetry breaking Fermi surface transformation. This is not easily reconciled with charge order correlations~\cite{CollignonPRB2021}. In the first place, it is difficult to envision how short-range charge correlations can impact the electronic structure and produce a Fermi surface transformation~\cite{MillisPRB2007}. 

The here reported high-resolution RIXS data on LSCO provide a different picture than existing literature that is mutually inconsistent. Our results resemble what has been reported in YBa$_2$Cu$_3$O$_{7-x}$ (YBCO).
High resolution RIXS measurements on YBCO and Nd$_{1+x}$Ba$_{2-x}$Cu$_3$O$_{7-\delta}$ reveal that the short-range correlations stem from quasi-elastic scattering processes~\cite{Arpaia906,WahlbergScience2021,YuPRX2020,ArpaiaJPSJ2021}. Also in LSCO,  it is not inconceivable that the short-range correlations result from dynamic charge fluctuations, for example due to electron-phonon coupling of a low-energy phonon branch. In fact, recent high-resolution O $K$-edge RIXS experiments on optimally doped LSCO indicate strong electron-phonon coupling at a low-energy phonon branch around $(h,0)=(1/4,0)$~\cite{HuangPRX2021}.
Just as in LSCO, the short and long-range charge correlations in YBCO and Nd$_{1+x}$Ba$_{2-x}$Cu$_3$O$_{7-\delta}$ occur with a very similar scattering vector. Furthermore, both systems seem to have a charge order quantum critical point near optimal doping. As long demonstrated, charge order in LSCO and YBCO differ by their respective ordering vectors $Q_{CO}\approx(1/4,0)$~\cite{tranquada} and $\approx(1/3,0)$~\cite{VinogradNatComm2021}. The fact that the short-range correlations are so aligned with the long-range ordering strongly suggests a direct link. If the short-range charge correlations should be understood from electron-phonon coupling, this could also be the case for long-range ordering. This raises the question as to why the La-214 and Y/Nd-1237 systems display different long-range incommensurabilities~\cite{BlackburnPRL2013}.
A possibility is that the detailed crystal structure generates different electron-phonon couplings~\cite{BanerjeeComPhy2020}. Recently, it was shown how electron-phonon coupling is enhanced inside the low-temperature tetragonal (LTT) crystal structure
of LESCO~\cite{WangSciAdv2021} --- suggesting that electron-phonon coupling assists in the formation of long-range stripe order.

We conclude by commenting on the fact that the charge order incommensurabilities in LSCO and YBCO have opposite doping dependencies. If electron-phonon coupling is solely responsible for charge ordering, a more universal picture would be expected. It is therefore conceivable that electron-electron coupling also plays an important role. In the La-based cuprates, many experimental results point to a strong coupling picture which would support the electron-electron interaction hypothesis. Irrespective of the responsible interactions, our observation of a charge order quantum critical point at optimally doped LSCO provides new perspectives. It opens for a unified picture of charge order as similar results are found in YBCO. The critical fluctuations around the critical point may also have bearing in superconductivity and strange metal properties.

\noindent\textbf{Methods}\\
High quality single crystals of LESCO and LSCO were grown using a floating zone method. These crystals have been used in previous studies~\cite{chang12,WangPRL2020,FrachetNatPhys2020,choi_disentangling_2020,IvashkoPRB2017, LaliberteNatComm2011}. All crystals were prealigned \textit{ex-situ} using x-ray LAUE backscattering and cleaved \textit{in-situ} by a standard top-post technique. RIXS experiments were carried out at the ADvanced RESonant Spectroscopies (ADRESS)~\cite{ghiringhelliREVSCIINS2006,strocovJSYNRAD2010} beamline of the Swiss Light Source (SLS) at the Paul Scherrer Institut and the ID32 RIXS beamline at the European Synchrotron Radiation Facility (ESRF)~\cite{BROOKES2018175}. 
%The scattering geometry is indicated in Fig.~\ref{fig:fig1}. 
Fixed angles of $\SI{130}{\degree}$ and $\SI{149.5}{\degree}$ between incident and scattered light was used for measurements at ADRESS and ID32, respectively. To determine zero energy loss, the low-energy part of each RIXS spectrum is analysed (see supplementary Fig. S1). Elastic scattering and a phonon excitation are fitted using Gaussian profiles where the width is set by the instrumental energy resolution. A damped harmonic oscillator response function is adopted to model the magnetic excitations and a quadratic function is added to mimic the background (BG).
Given the quasi-two-dimensional character of this system, we consider only the in-plane momentum transfer which can be controlled by varying the incident angle $\theta$ and sample azimuthal angle $\phi$. Wave vector $\mathbf{Q}$ at $(q_{x},q_{y},q_{z})$ is defined as $(h,k,\ell)=(q_{x}a/2\pi,q_{y}b/2\pi,q_{z}c/2\pi)$ reciprocal lattice units (r.l.u.) using pseudo-tetragonal notation, with $a\approx b\approx3.79$~\AA~and $c\approx13.1$~\AA. % ~for LESCO.
Energy resolution, expressed in standard Gaussian deviation ranges from $\sigma_{\rm{_G}}\approx 49$ to $57$~meV for ADRESS data and $\sigma_{\rm{_G}}\approx 15$~meV for the ID32 data. Across all experiments, base temperature was in the range of $\sim 15 - 25$ K.
 
\noindent\textbf{Acknowledgements}\\
We thank D. LeBeouf and M.-H. Julien for discussions and comments to the manuscript. Part of the experiments have been performed at the ADRESS beamline of the Swiss Light Source at the Paul Scherrer Institut (PSI). We acknowledge Eugenio Paris for support at the beamline. K.v.A., Q.W., M.H., T.S., Y.T, and J.C. acknowledge support by the Swiss National Science Foundation through Grant Numbers BSSGI0-155873, 200021-188564, CRSII2\textunderscore160765\slash1 and 200021-178867.  K.v.A. is grateful for the support from the FAN Research Talent Development Fund - UZH Alumni. K.v.A. and Q.W. thank the Forschungskredit of the University of Zurich, grant no. [FK-21-105] and [FK-20-128]. Y.S. thanks the Chalmers Area of Advances-Materials Science and the Swedish Research Council (VR) with a starting Grant (Dnr. 2017-05078) for funding. T.C.A. acknowledges funding from the European Union’s Horizon 2020 research and innovation programme under the Marie Skłodowska-Curie grant agreement No. 701647 (PSI-FELLOW-II-3i program).
\\
 
\vspace{2mm}
\noindent\textbf{Authors contributions}\\
T.K., N.M., M.O., and E. P. grew the LSCO single crystals whereas S.P., T.T., and H.T.  synthesized LESCO single crystals. Q.W., K.v.A., S.M., M.H., D.M., N.B.B., D.B., W.Z., T.C.A., Y.T., Y.S., D.G.M., T.S., and J.C. carried out the RIXS experiments. K.v.A analysed the data with support from Q.W. and J.C..
K.v.A. and J.C. wrote the manuscript with input from all authors.

\vspace{2mm}
\noindent\textbf{Competing interests}\\
The authors declare no competing interests.

\vspace{2mm}
\noindent\textbf{Data and materials availability}\\ 
The data that support the findings of this study are available from the corresponding author upon request.
   
\bibliography{Eu-LSCO_RIXS}

\onecolumngrid
\newcommand{\beginsupplement}{
        \setcounter{table}{0}
        \renewcommand{\thetable}{S\arabic{table}}
        \setcounter{figure}{0}
        \renewcommand{\figurename}{\textbf{Figure S}}}

\beginsupplement
\clearpage
\begin{center}
{\Large Supplemental Information\\[4mm]
\large Fate of charge order in overdoped La-based cuprates}    
\end{center}

\begin{figure}[h]
\centering
\includegraphics[width=0.8\textwidth]{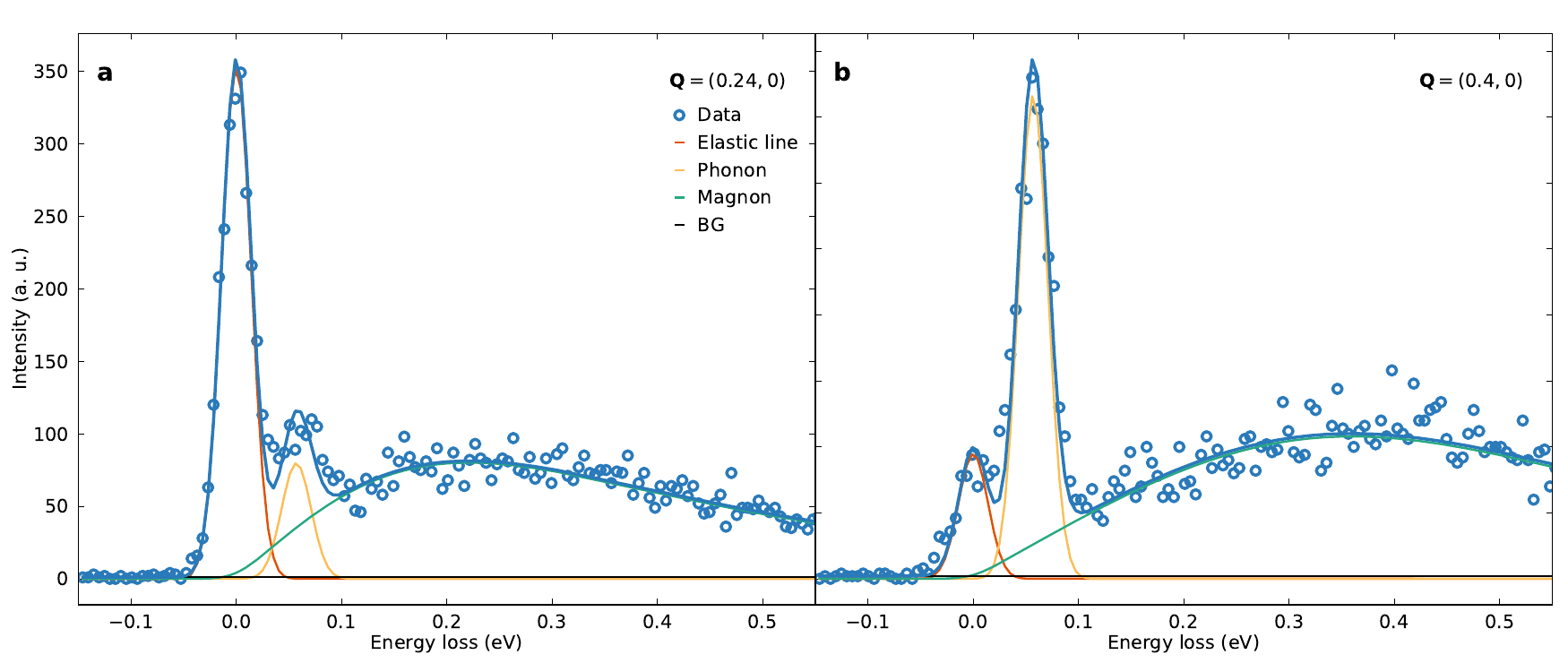}
\caption{\textbf{Raw RIXS spectra of LSCO $\mathbf{x=0.145}$ and associated fits}. Representative raw RIXS spectra at (a) the charge order wave vector and (b) away from charge order recorded at 37~K. Solid lines indicate fits of respective elastic, phonon, magnon and background (BG) components of the spectra.}
\label{fig:S1}
\end{figure}
\clearpage

\begin{figure}
\centering
\includegraphics[width=0.8\textwidth]{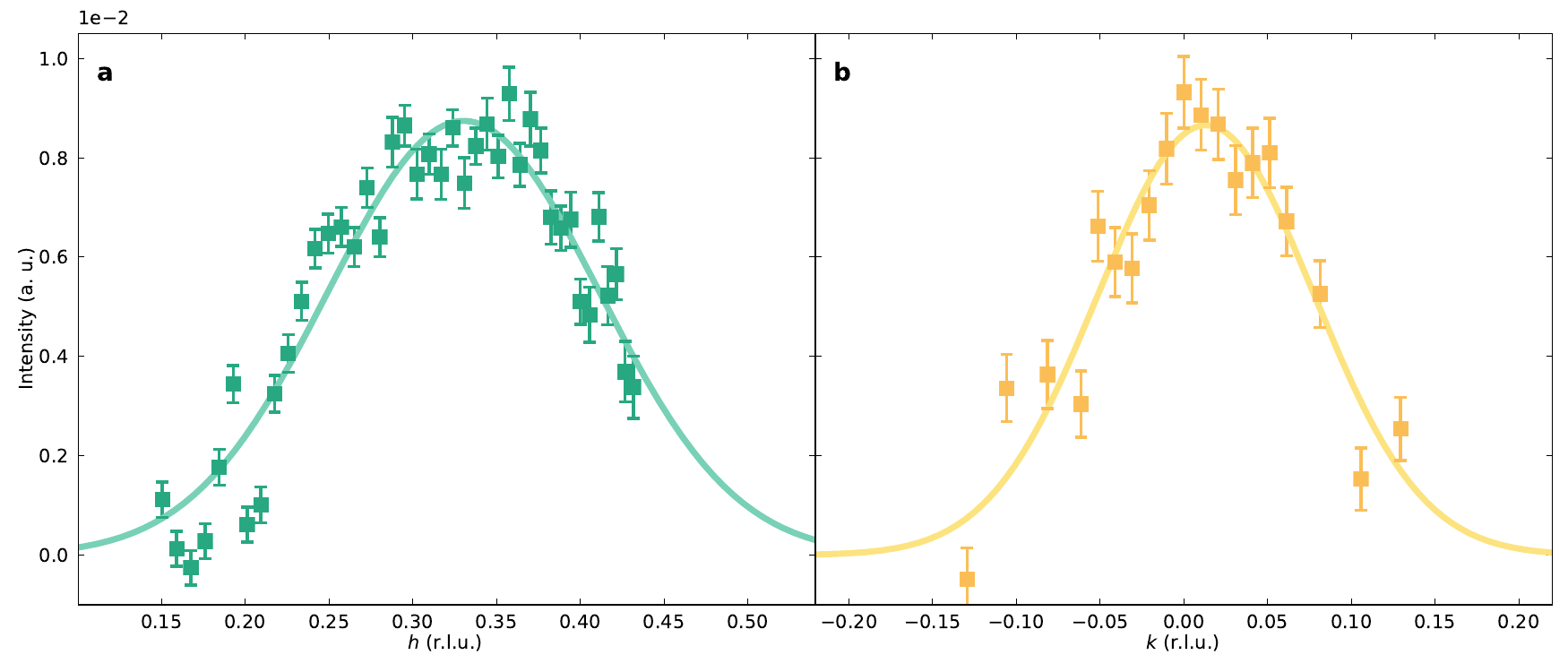}
\caption{\textbf{Elastic scattering recorded on LESCO $\mathbf{x=0.21}$}. Longitudinal (a) and transverse (b) elastic scans through the charge order wave vector. Data was recorded with medium resolution at base temperature. Solid lines are Gaussian fits.}
\label{fig:S2}
\end{figure}
\clearpage

\begin{figure}
\centering
\includegraphics[width=0.8\textwidth]{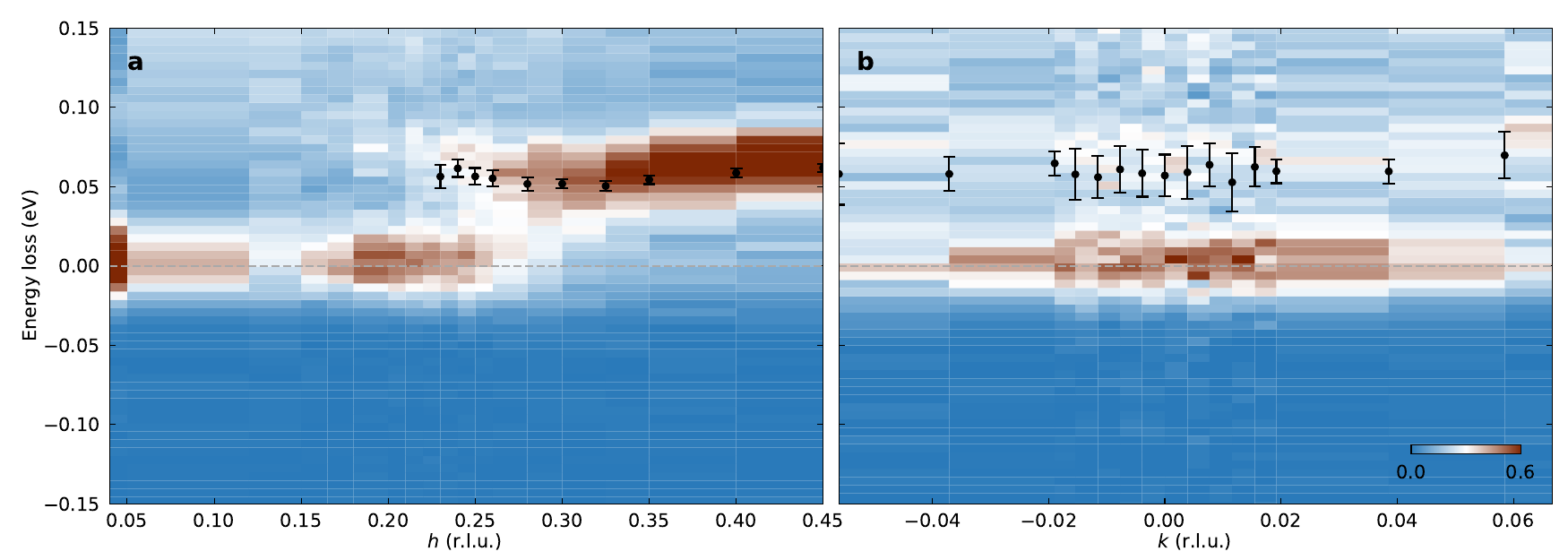}
\caption{\textbf{RIXS intensity recorded on LSCO $\mathbf{x=0.16}$}. (a) RIXS spectra probed in longitudinal $h$ and (b) transverse $k$ directions on LSCO $x = 0.16$. Data are recorded at a high resolution beamline with a total energy resolution of 33 meV and at a temperature of 35 K. Horizontal dashed lines mark the zero energy loss. The black circles mark the phonon dispersion determined from fitting the spectra.}
\label{fig:S3}
\end{figure}

\end{document}